\documentclass[prl,twocolumn,superscriptaddress,showpacs,a4paper]{revtex4}

\usepackage[pdftex]{graphicx}
\usepackage{booktabs}
\usepackage{pdflscape}
\usepackage{bm}        
\usepackage{amssymb} 
\usepackage{color}

\newcommand{\Ham}{\mathcal{H}}     
\newcommand{\spin}[1]{\sigma_{#1}} 
\newcommand{\nn}[2]{\left<#1,#2\right>} 
\newcommand{\nnn}[2]{\left<\left<#1,#2\right>\right>} 
\newcommand{\plaq}[4]{\left[#1,#2,#3,#4\right]} 
\newcommand{\nnsum}{\sum\limits_{\nn{i}{j}}\spin{i}\spin{j}} 
\newcommand{\nnnsum}{\sum\limits_{\nnn{i}{j}}\spin{i}\spin{j}} 
\newcommand{\plaqsum}{
  \sum\limits_{\plaq{i}{j}{k}{l}}\spin{i}\spin{j}\spin{k}\spin{l}}

\begin{document}


\title{Non-{Standard} Finite-Size Scaling at First-Order Phase Transitions}

\date{March 2014}

\author{Marco Mueller}
\affiliation{Institut f\"ur Theoretische Physik, Universit\"at Leipzig,
   Postfach 100\,920, D-04009 Leipzig, Germany}

\author{Wolfhard Janke}
\affiliation{Institut f\"ur Theoretische Physik, Universit\"at Leipzig,
   Postfach 100\,920, D-04009 Leipzig, Germany}

\author{Desmond A.\ Johnston}
\affiliation{Department of Mathematics, 
School of Mathematical and Computer Sciences,
Heriot-Watt University, Riccarton, Edinburgh EH14 4AS, Scotland}

\begin{abstract}

We note that the standard inverse system volume scaling  for finite-size corrections at a first-order phase transition (i.e., $1/L^3$ for an $L \times L \times L$ lattice in $3D$)  is transmuted to $1/L^2$ scaling if there is an exponential low-temperature phase degeneracy. The gonihedric Ising model which has a four-spin interaction, plaquette Hamiltonian provides an exemplar of just such a system. We use multicanonical simulations of this model to generate high-precision data which  provides strong confirmation of the non-{standard} finite-size scaling law. The dual to the gonihedric model, which is an anisotropically coupled Ashkin-Teller model, has a similar degeneracy and also  displays the non-{standard} scaling. 
\end{abstract}


 
\maketitle


First-order phase transitions are ubiquitous in nature \cite{ubiquity}.
Pioneering studies of finite-size scaling for first-order transitions were carried out in \cite{pioneer}
and subsequently pursued in detail in \cite{furtherd}. Rigorous results for periodic boundary conditions were further derived in \cite{rigorous,fsspbc}.
It is possible to go quite a long way in discussing the scaling laws for such first-order
transitions using a simple heuristic two-phase model~\cite{twophasemodel}. We assume that a
system spends a fraction $W_{\rm o}$ of the total time in one of the $q$
ordered phases and a fraction $W_{\rm d} = 1 - W_{\rm o}$ in the disordered
phase with corresponding energies $\hat{e}_{\rm o}$ and $\hat{e}_{\rm d}$,
respectively. The hat is introduced for quantities evaluated at the inverse
transition temperature of the infinite system, $\beta^\infty$.
Neglecting all fluctuations within the phases and treating the phase transition
as a sharp jump between the phases, the energy moments become $\left<e^n\right> =
W_{\rm o}\hat{e}_{\rm o}^n + (1-W_{\rm o})\hat{e}_{\rm d}^n$. The specific heat
$C_V(\beta, L) = -\beta^2\partial e(\beta, L)/\partial\beta$ then reads
\begin{equation}
  C_V(\beta, L) = L^d\beta^2\left(\left< e^2\right> - \left< e \right>^2\right) =
  L^d\beta^2 W_{\rm o}(1-W_{\rm o})\Delta \hat{e}^2
  \label{eq:specheat}
\end{equation}
with $\Delta \hat{e} = \hat{e}_{\rm d} - \hat{e}_{\rm o}$. It has a maximum
$C_V^{\rm max} = L^d (\beta^\infty\Delta \hat{e}/2)^2$ 
at $\beta^{C_V^{\rm max}}(L)$ for $W_{\rm o} = W_{\rm d}
= 0.5$, i.e., where the disordered and
ordered peaks of the energy probability density have equal weight.  The
probability of being in any of the ordered states or the disordered state is
related to the free energy densities $\hat{f}_{\rm o}, \hat{f}_{\rm d}$ of the states,
\begin{equation}   
p_{\rm o}\propto e^{-\beta L^d \hat{f}_{\rm o}}\mbox{ and } p_{\rm d} \propto
  e^{-\beta L^d \hat{f}_{\rm d}} \; ,
\end{equation}
and by construction the fraction of time spent in the ordered states must be
proportional to $q p_{\rm o}$. Thus for the ratio of fractions we find $W_{\rm
o}/W_{\rm d} \simeq q e^{- L^d \beta \hat{f}_{\rm o}}/ e^{-\beta L^d \hat{f}_{\rm d}}$ (up to exponentially small corrections in $L$ \cite{rigorous,fsspbc,borgs-janke,twophasemodel}).
Taking the logarithm of this ratio gives $\ln (W_{\rm o}/W_{\rm d}) \simeq \ln q + L^d\beta(\hat{f}_{\rm d}
- \hat{f}_{\rm o})$. At the specific-heat maximum $W_{\rm o} = W_{\rm d}$, so
we find by an expansion around $\beta^\infty$
\begin{equation}
   0 = \ln q + L^d\Delta \hat{e}(\beta -
  \beta^\infty) + \dots \end{equation}
which can be solved for the finite-size peak location of the specific heat:
\begin{equation}
  \beta^{C_V^{\rm max}}(L) = \beta^\infty - \frac{\ln q}{L^{d}\Delta\hat{e}} +
  \dots
  \label{eq:fss:beta:specheat}
\end{equation}
Although this is a rather simple toy model, it is known to capture the
essential features of first-order phase transitions and  to correctly predict the
prefactors of the leading finite-size scaling corrections for a class of models
with a contour representation, such as the $q$-state Potts model, where a
rigorous theory also exists~\cite{fsspbc}. Similar calculations give~\cite{twophasemodel,lee-kosterlitz}
\begin{equation}
  \beta^{B^{\rm min}}(L) = \beta^\infty - \frac{\ln(q\hat{e}^2_{\rm
  o}/\hat{e}^2_{\rm d})}{L^d\Delta \hat{e}}  + \dots 
  \label{eq:fss:beta:binder}
\end{equation}
for the location $\beta^{B^{\rm min}}(L)$ of the minimum of the energetic
Binder parameter 
\begin{equation}
  B(\beta, L) = 1 - \frac{\langle e^4\rangle}{3\langle e^2 \rangle^2} \; .
  \label{eq:binder}
\end{equation}

Normally the degeneracy $q$ of the low-temperature phase does not change with system size  and the generic finite-size scaling
behaviour of a first-order transition thus has a leading contribution proportional
to the inverse volume $L^{-d}$. We can see from Eqs.~(\ref{eq:fss:beta:specheat}), (\ref{eq:fss:beta:binder}) that if the degeneracy $q$ of the low-temperature phase depends exponentially on the system size, $q \propto e^L$, this would be modified. One  model with precisely this feature is a $3D$ plaquette (4-spin) interaction Ising model on a cubic lattice where $q=2^{3L}$ on an $L^3$ lattice \cite{degen}.
This is a member of a family of so-called gonihedric Ising models \cite{goni1}
whose Hamiltonians contain, in general,
nearest $\nn{i}{j}$, next-to-nearest $\nnn{i}{j}$ and plaquette interactions
$\plaq{i}{j}{k}{l}$. {These were originally formulated as a lattice discretization of string-theory actions in {high-energy} physics  which depend solely on the extrinsic curvature of the string worldsheet \cite{goni2}}.

The weights of the different interactions are fine-tuned
so that the area of spin-cluster boundaries does not contribute to the partition
function. However, edges and self-intersections of spin-cluster boundaries {\it are}
weighted, leading to 
\begin{equation}
  \Ham^{\kappa} = -2\kappa\nnsum+\frac{\kappa}{2}\nnnsum-\frac{1-\kappa}{2}\plaqsum
  \label{eq:ham:goni}
\end{equation}
where $\kappa$ effectively parametrizes the  self-avoidance of the spin-cluster boundaries.
The purely plaquette Hamiltonian  with $\kappa=0$ that we consider here,
\begin{equation}
  \Ham = -\frac{1}{2}\plaqsum \;,
  \label{eq:ham:gonikappa0}
\end{equation}
allows spin-cluster boundaries to intersect without energetic penalty. It has
attracted {particular  attention in its own right}, since it displays a strong first-order transition
\cite{firstorder}
and evidence of glass-like behaviour at low temperatures \cite{glassy}. 
{Computer simulation studies of this model were plagued, 
however, by enduring inconsistencies in the estimates of the transition 
temperature.}
The dual to this plaquette gonihedric Hamiltonian
can be written as an anisotropic Ashkin-Teller model \cite{dual} in which
two spins $\sigma, \tau$ live on each
vertex, with nearest-neighbour interactions along the $x,y,$
and $z$-axes,
\begin{equation} 
  \Ham^{\mathrm{d}} = -\frac{1}{2}\sum\limits_{\nn{i}{j}_x}\sigma_i\sigma_j -
  \frac{1}{2}\sum\limits_{\nn{i}{j}_y}\tau_i\tau_j
  -\frac{1}{2}\sum\limits_{\nn{i}{j}_z}\sigma_i\sigma_j\tau_i\tau_j\;, 
  \label{eq:ham:dual}
\end{equation}
and this too has an exponentially degenerate ground state. We assume that the 
exponential degeneracy also  extends into the low-temperature phase for 
the dual model and check the consistency of the assumption in the numerical 
scaling analysis below.   

In the gonihedric model with $q=2^{3L}$ Eqs.~(\ref{eq:fss:beta:specheat}), (\ref{eq:fss:beta:binder})
become
\begin{eqnarray} \beta^{C_V^{\rm max}}(L) &=& \beta^\infty - \frac{\ln
    2^{3L}}{L^3\Delta\hat{e}} \nonumber\\
    &+& {\cal O}\left( (\ln 2^{3L})^2 L^{-6} \right)
    \label{eq:fss:beta:specheat2} \\
    &=& \beta^\infty - \frac{3\ln 2}{L^{2}\Delta\hat{e}} + {\cal O}\left( L^{-4}
    \right)\nonumber
\end{eqnarray} 
and
\begin{eqnarray} 
  \beta^{B^{\rm min}}(L) &=& \beta^\infty - \frac{\ln(2^{3L}\hat{e}^2_{\rm
  o}/\hat{e}^2_{\rm d})}{L^3\Delta \hat{e}}\nonumber\\  
  &+& {\cal O}\left(
  (\ln(2^{3L}\hat{e}^2_{\rm o}/\hat{e}^2_{\rm d}))^2 L^{-6}\right)
  \label{eq:fss:beta:binder2} \\ &=&
  \beta^\infty - \frac{3\ln2}{L^2\Delta \hat{e}}  - \frac{\ln(\hat{e}^2_{\rm
  o}/\hat{e}^2_{\rm d})}{L^3\Delta\hat{e}} + {\cal O}\left( L^{-4}\right)\nonumber
\end{eqnarray} 
and the leading contribution to the finite-size corrections is now $\propto
L^{-2}$.  For the extremal values one expects 
\begin{equation} 
  C_V^{\rm max}(L) = L^3\left( \frac{\beta^\infty \Delta\hat{e}}{2}\right)^2 +
  {\cal O}(L)
  \label{eq:fss:specheat} 
\end{equation} 
and 
\begin{equation} B^{\rm
  min}(L) = 1 - \frac{1}{12}\left( \frac{\hat{e_{\rm o}}}{\hat{e_{\rm d}}} +
  \frac{\hat{e_{\rm d}}}{\hat{e_{\rm o}}} \right)^2  + {\cal O}(L^{-2}) \; .
  \label{eq:fss:binder} 
\end{equation}

The inverse temperature where both peaks of the energy probability density have
equal {\it weight}, $\beta^{\rm eqw}(L)$, has a behaviour that coincides with the
scaling of the location of the specific-heat maximum in Eq.~(\ref{eq:fss:beta:specheat2}),
\begin{equation} 
  \beta^{\rm eqw}(L) = \beta^\infty - \frac{3\ln 2}{L^{2}\Delta\hat{e}} + {\cal
  O}\left( L^{-4} \right) \; .  
  \label{eq:fss:beta:eqw} 
\end{equation} 
It can be shown that even the ${\cal
  O}\left( L^{-4} \right)$ terms in Eqs.~(\ref{eq:fss:beta:specheat2}), (\ref{eq:fss:beta:binder2}), and (\ref{eq:fss:beta:eqw}) coincide exactly \cite{highero}.
The leading term in the  scaling behaviour of the inverse temperature of equal peak
{\it height}, $\beta^{\rm eqh}(L)$, is also of the form (\ref{eq:fss:beta:eqw}) 
but similar to $\beta^{B^{\rm min}}(L)$ the higher order corrections start already 
with ${\cal O}\left( L^{-3} \right)$ and are different.

\begin{figure}[tpb] 
  \begin{center} 
    \includegraphics[scale=0.7]{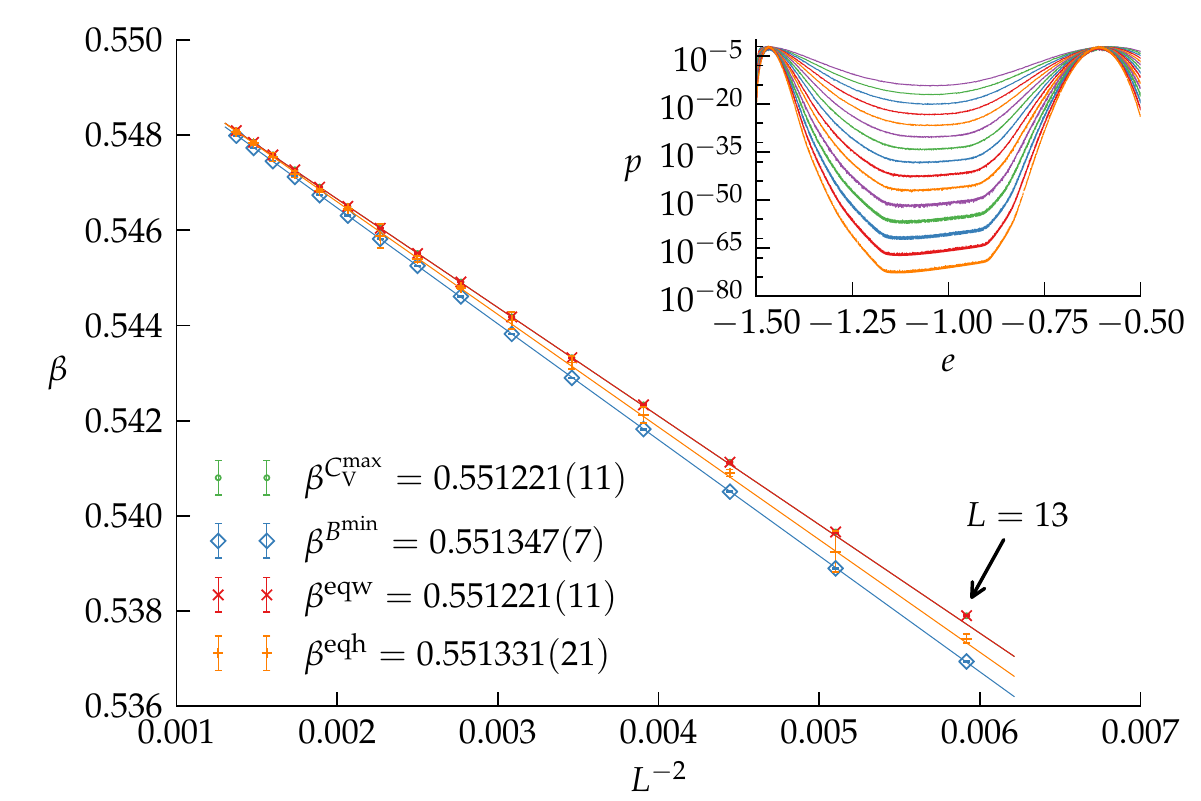} 
    \caption{
      Best fits using the leading $1/L^2$ scaling obtained for the original 
      model (\ref{eq:ham:gonikappa0})  
      using the (finite lattice) peak locations for  the specific
      heat $C_V^{\rm max}$, Binder's energy cumulant $B^{\rm min}$; or inverse
      temperatures $\beta^{\rm eqw}$ and $\beta^{\rm eqh}$, where the two peaks
      of the energy probability density are of same weight or have equal
      height, respectively. The values for $\beta^{\rm eqw}$ and $\beta^{C_V^{\rm
      max}}$ are indistinguishable in the plot. The higher order corrections which we discuss in detail
      in \cite{highero} give the slightly different slopes.
      The inset shows the energy
      probability density $p(e)$ over $e = E/L^d$ at $\beta^{\rm eqh}$ for
      lattices with linear length $L\in \left\{ 13, 14, \dots, 26, 27 \right\}$. }
    \label{fig:fit:orig-pbc} 
  \end{center} 
\end{figure}

To overcome supercritical slowing down near
first-order phase transitions where canonical simulations tend to get trapped
in one phase and evade other problems such as hysteresis, we employed the multicanonical
Monte Carlo algorithm~\cite{muca}. Our approach is to systematically improve guesses of the energy
probability distribution 
{using recursive estimates~\cite{mucaweights} before the actual 
production run with of the order of $(100 - 1000) \times 10^6$ sweeps {for the original model and $4 \times 10^6$ sweeps for its dual.}}
Rare states lying between the ordered and disordered phases are then promoted
artificially, decreasing the autocorrelation time and allowing the system to
oscillate more rapidly between phases. 
{
For the original model (\ref{eq:ham:gonikappa0}), we took measurements
only every $V=L^3$ sweeps to reduce the autocorrelation time $\tau_{\rm
int}^{\rm meas}$ in the actual time series of the measurements.  Simulations
were terminated after approximately $500$ hours of real time for each lattice
individually. We therefore collected less statistics for larger lattices. 
{Still,} the largest lattice of $27^3$ spins effectively transited
more than 250 times between the two phases during the simulation, even though rare
states are suppressed by more than 60 orders of magnitude compared to the most probable
states (see the inset in Fig.~\ref{fig:fit:orig-pbc}). For the dual model, we
took measurements every sweep, therefore the autocorrelation time $\tau_{\rm
int}^{\rm meas}$ 
is much larger here.}
Canonical estimators can then be
retrieved by weighting the multicanonical data to yield Boltzmann-distributed
energies. Reweighting techniques are very powerful when combined with multicanonical
simulations, and allow the calculation of observables over a broad range of
temperatures.
{
Errors on the measured quantities have been extracted by jackknife 
analysis~{\cite{jackknife}} using $20$
blocks for each lattice size.}

Standard observables such as the specific heat (\ref{eq:specheat}) and
Binder's energy cumulant (\ref{eq:binder}) have been calculated 
{from our data as function of temperature by reweighting}
for both the gonihedric Ising
model in Eq.~(\ref{eq:ham:gonikappa0}) and its dual in Eq.~(\ref{eq:ham:dual}). 
{This enables us to determine the}
positions of their peaks, $\beta^{C_V^{\rm max}}(L)$ and $\beta^{B^{\rm
min}}(L)$, 
{with high precision.} 
To obtain the other quantities of interest, $\beta^{\rm eqw}(L)$,
 and $\beta^{\rm eqh}(L)$, we use reweighting techniques to get an estimator of the energy
probability densities $p(e)$ at different temperatures. $\beta^{\rm eqw}$ is chosen
systematically to minimize 
\begin{equation} 
  D^{\rm eqw}(\beta) = \left( \sum_{e < e_{\rm min}} p(e, \beta) - \sum_{e \geq
  e_{\rm min}} p(e, \beta) \right)^2 
\end{equation}
where the energy of the minimum between the two peaks, $e_{\rm min}$, is
determined beforehand to distinguish between the different phases.
Similarly, $\beta^{\rm eqh}$ is chosen to minimize 
\begin{equation} 
  D^{\rm eqh}(\beta) =  \left( \max_{e < e_{\rm min}}\{p(e, \beta)\} - \max_{e
  \geq e_{\rm min}}\{p(e, \beta)\} \right)^2
\end{equation} 
as function of $\beta$.

\begin{figure}[htpb] 
  \begin{center} 
    \includegraphics[scale=0.7]{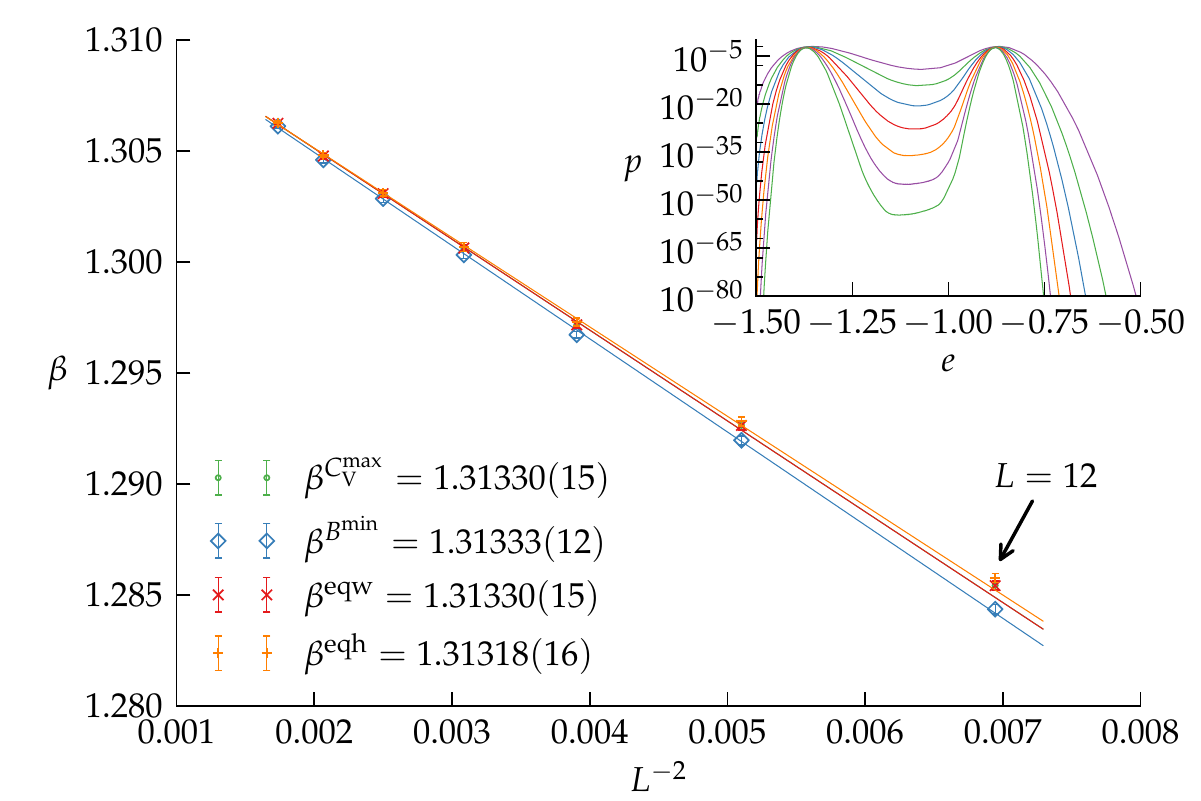} 
    \caption{Same as Fig.~\ref{fig:fit:orig-pbc} but for the dual 
    model (\ref{eq:ham:dual}).
    Here the inset shows the energy
    probability density $p(e)$ at $\beta^{\rm eqh}$ for
    lattices with linear length $L\in \left\{ 12, 14, \dots, 22, 24
    \right\}$.} 
    \label{fig:fit:dual} 
  \end{center} 
\end{figure} 

\begin{figure}[htpb] 
  \begin{center} 
    \includegraphics[scale=0.7]{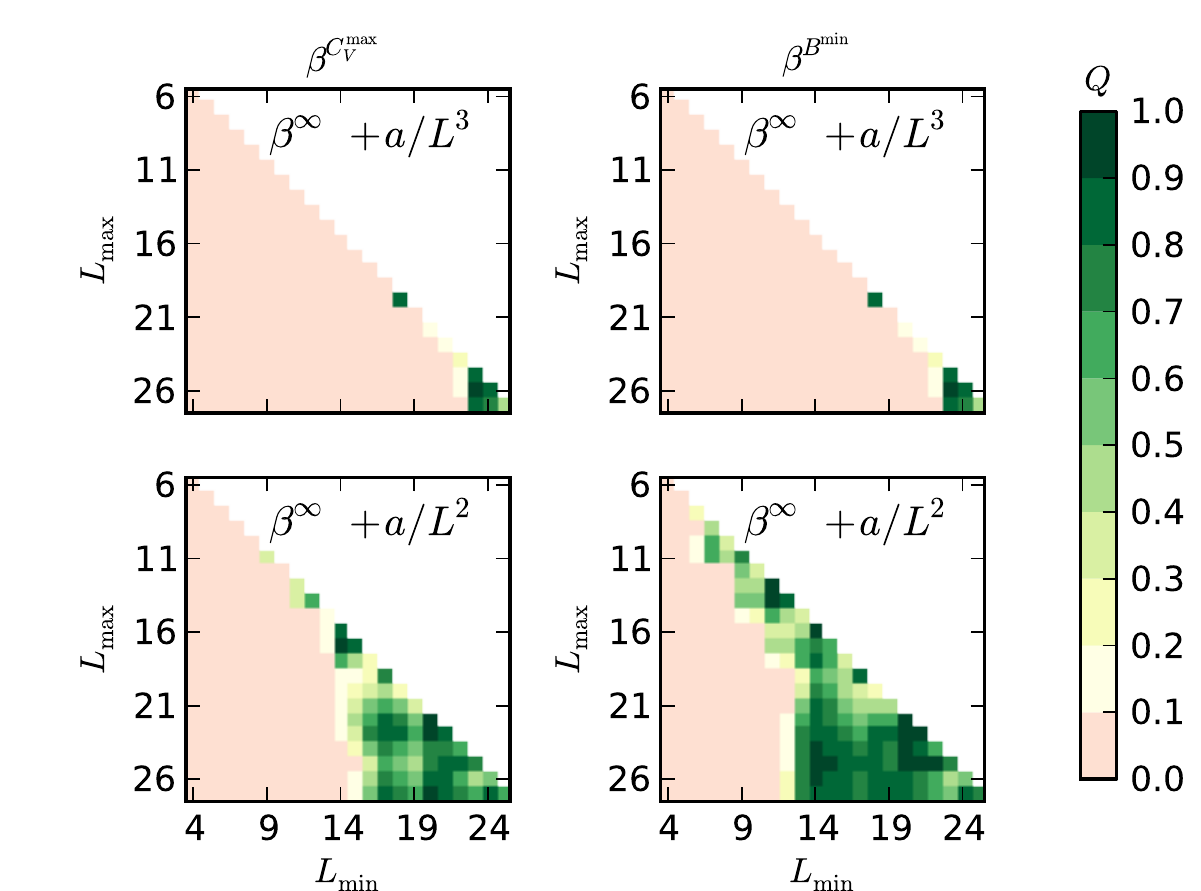}
    \caption{Plot of the goodness-of-fit parameter $Q$ for fits on the
      extremal locations of the specific heat, $\beta^{C_V^{\rm max}}$, and
      Binder's energy cumulant, $\beta^{B^{\rm min}}$ of the original model
      for different fitting ranges
      \mbox{$L_{\rm min}$ -- $L_{\rm max}$}. Upper row: Standard ($1/L^3$) finite-size
      scaling ansatz. Lower row: Transmuted ($1/L^2$) finite-size scaling.} 
      \label{fig:fitting_quality} 
  \end{center}
\end{figure}

{
The data and fits for the inverse transition temperatures are shown in 
Fig.~\ref{fig:fit:orig-pbc} for the original and in Fig.~\ref{fig:fit:dual} for
the dual model. Unusually, the estimates for $\beta^{C_V^{\rm max}}$ and
$\beta^{\rm eqw}$ fall together because of the aforementioned equality of the 
${\cal O} (L^{-4})$ corrections in the scaling ansatz for these quantities in
Eqs.~(\ref{eq:fss:beta:specheat2}) and (\ref{eq:fss:beta:eqw}). The fits}
have been carried out according to the non-{standard} scaling laws with
$1/L^2$ corrections. We have
left out the smaller lattices systematically, until a goodness-of-fit value of
at least $Q = 0.5$ was found for each observable individually. 
{
From error weighted averages (refraining from a full cross-correlation analysis~\cite{combinefitsweigel})
of the inverse transition temperatures $\beta^{C_V^{\rm max}}, 
\beta^{B^{\rm min}}, \beta^{\rm eqw}$, and $\beta^{\rm eqh}$ given in 
Figs.~\ref{fig:fit:orig-pbc} and \ref{fig:fit:dual} we find}
\begin{eqnarray*}
	\beta^{\infty} &=& {0.551\,291(7)} , \\
\beta^{\infty}_{\text{dual}} &=& 1.313\,29(12)
\end{eqnarray*} 
for the infinite lattice inverse transition temperatures of the original and dual models, where
the final error estimates are taken as the smallest error bar of the contributing $\beta$ estimates. 

The temperature $\beta_{\text{dual}}^\infty$ of the dual model is related to the temperature in
the original model, $\beta^\infty$, by the 
duality transformation 
\begin{equation}
  \beta^\infty = -\ln\left( \tanh\left( \frac{\beta_{\text{dual}}^\infty}{2}
  \right) \right) \;. \label{eq:duality} 
  \end{equation} 
Applying standard
error propagation, we retrieve a value of $\beta^{\infty} =
  0.551\,43(7)$ for the
original model from dualizing $\beta^{\infty}_{\text{dual}} = 1.313\,29(12)$. The estimated values of the critical temperature from the direct simulation and the simulation of
the dual model are  thus in good agreement, considering that higher order and exponential
corrections~\mbox{\cite{rigorous,fsspbc,borgs-janke,twophasemodel}} in the finite-size
scaling were not included. {The application of the non-standard
finite-size scaling laws thus settles the enduring and very puzzling inconsistencies
in previous estimates of the transition temperature for these models.}

The great precision of our simulation results and the broad range of lattice
sizes clearly excludes fits to the standard finite-size scaling ansatz,
where the first correction is proportional to the inverse volume. This is
demonstrated in Fig.~\ref{fig:fitting_quality}, where the upper two plots
show how the standard finite-size scaling ansatz leads to poor results.
Acceptable fits are only achieved for a narrow fitting range with almost no
degrees of freedom left.  The  fits using the
non-{standard} laws shown in the lower plots are much better over a
broad range of fitting intervals.  Fits carried out using the traditional
inverse volume scaling ansatz led to an inverse transition temperature of 
{$0.549\,987(30)$}
for the
original model (with only 
{$5$}
degrees of freedom left per fit and 
{$Q \approx 0.8$)}
and
$1.310\,29(19)$ for the dual model (best fits with $2$ degrees of freedom and 
{$Q \approx 0.3$} for all fits), which translates to $0.553\,17(11)$ using the duality
relation (\ref{eq:duality}). These values are about $30$ error bars apart.
Since the dual model clearly displays the non-{standard} scaling behaviour this confirms our initial assumption  that the low-temperature phase (and not just the ground state) is exponentially degenerate in this case also.

We should  emphasize that the considerations described here for the gonihedric model and its dual apply generically to any 
models which  have a low-temperature phase degeneracy
which depends exponentially on the system size. {
Apart from higher-dimensional variants of the gonihedric model \cite{higherdG}, there are numerous other fields where the scenario could be realized. Examples range from ANNNI models \cite{selke} to spin ice systems \cite{gingras} and topological {``orbital''} models in the context of quantum computing \cite{nussinov} which all share an extensive ground-state degeneracy. It would be worthwhile to explore to what extent these ground states evolve as stable 
low-temperature phases and whether they eventually undergo a first-order transition into the disordered phase with increasing temperature. Among the orbital models for transition metal compounds, a particularly promising candidate is the three-dimensional classical compass or $t_{2g}$ orbital model  \cite{compass0} where a highly degenerate ground state is well known and signatures of a first-order transition into the disordered phase have recently been found numerically \cite{compass}. This has a 
{ground-state}
degeneracy of $2^{3 L^2}$ so non-standard scaling corrections, this time of $O(1/L)$,
might be expected at its first-order transition point with periodic boundary conditions.}

This work was supported by
the Deutsch-Franz\"osische Hochschule (DFH-UFA) under Grant No.\ CDFA-02-07.



\end{document}